\documentclass[12pt]{article}
\usepackage{times}
\usepackage{geometry}
\usepackage[utf8]{inputenc}
\usepackage{enumitem,amssymb}
\usepackage{ragged2e}
\usepackage{wrapfig}
\newlist{thematic}{itemize}{8}
\setlist[thematic]{label=$\square$}
\usepackage{pifont}

\usepackage{verbatim}
\usepackage{natbib}
\usepackage{mathtools}
\usepackage{hyperref}

\newcommand{\apj}{ApJ}

\newcommand{\apjs}{ApJS}

\newcommand{\mnras}{MNRAS}
\newcommand{\nat}{Nature}

\newcommand{\pasp}{PASP}

\newcommand{\aap}{A\&A}

\newcommand\procspie{Proc.~SPIE}

\usepackage{color}
\usepackage{xcolor}

\definecolor{goodblue}{HTML}{4575B4}
\newcommand{\goodblue}{\color{goodblue}}
\definecolor{goodred}{HTML}{D73027}

\begin{document}
\thispagestyle{empty}
\raggedright
\huge
Astro2020 APC White Paper\\ 
\vspace*{10pt}
On the need for synthetic data and robust data simulators in the 2020s  \linebreak
\normalsize

\noindent \textbf{Thematic Areas:}\  \hspace*{4pt} $\square$ Ground-Based Project \hspace*{4pt} $\square$ Space-Based Project \hspace*{4pt}
$\square$ Infrastructure Activity \hspace*{4pt}\\
$\boxtimes$ Technological Development Activity \hspace*{4pt}
$\square$ State of the Profession Consideration\linebreak

\textbf{Principal Author:}

Name: Molly S.\ Peeples
 \linebreak						
Institution: Space Telescope Science Institute / Johns Hopkins University
 \linebreak
Email: \texttt{molly@stsci.edu}
 \linebreak
Phone: 1 (410) 338-2451
 \linebreak

\textbf{Co-authors:}\\
Bjorn Emonts, National Radio Astronomy Observatory, \texttt{bemonts@nrao.edu}\\
Mark Kyprianou, Space Telescope Science Institute, \texttt{kyp@stsci.edu}\\
Matthew T.\ Penny, The Ohio State University, \texttt{penny@astronomy.ohio-state.edu}\\ 
Gregory F.\ Snyder, Space Telescope Science Institute, \texttt{gsnyder@stsci.edu}\\
Christopher C. Stark, Space Telescope Science Institute, \texttt{cstark@stsci.edu}\\
Michael Troxel, Duke University, \texttt{michael.troxel@duke.edu}\\
Neil T.\ Zimmerman, NASA Goddard Space Flight Center, \texttt{neil.t.zimmerman@nasa.gov}\\
\hangindent=0.7cm
John ZuHone, Harvard-Smithsonian Center for Astrophysics, \texttt{john.zuhone@cfa.harvard.edu}
 \linebreak

\textbf{Endorsers:}\\
Chuanfei Dong, Princeton University, 
\texttt{dcfy@princeton.edu}\\
Nimish Hathi, Space Telescope Science Institute,
\texttt{nhathi@stsci.edu}\\
Andrew Hearin, Argonne National Lab,
\texttt{ahearin@anl.gov}\\
Sangeeta Malhotra, NASA Goddard Space Flight Center, \texttt{sangeeta.malhotra@nasa.gov}\\
Raymond C. Simons, Space Telescope Science Institute,
\texttt{rsimons@stsci.edu}\\

\vspace{10pt}

\justifying
\noindent\textbf{Abstract:}
\noindent  

As observational datasets become larger and more complex, so too are the questions being asked of these data. Data simulations, i.e., synthetic data with properties (pixelization, noise, PSF, artifacts, etc.) akin to real data, are therefore increasingly required for several purposes, including: (1) testing complicated measurement methods, (2) comparing models and astrophysical simulations to observations in a manner that requires as few assumptions about the data as possible, (3) predicting observational results based on models and astrophysical simulations for, e.g., proposal planning, and (4) mitigating risk for future observatories and missions by effectively priming and testing pipelines. We advocate for an increase in using synthetic data to plan for and interpret real observations as a matter of routine. This will require funding for (1) facilities to provide robust data simulators for their instruments, telescopes, and surveys, and (2) making synthetic data publicly available in archives (much like real data) so as to lower the barrier of entry to all. 
\thispagestyle{empty}

\pagebreak
\setcounter{page}{1} 
As observational datasets become larger and more complex, so too are the questions being asked of these data. Increasingly, asking and answering these questions requires analyzing data simulations, i.e., synthetic data with properties (pixelization, noise, PSF, artifacts, etc.) akin to real data. In the past the assumptions going into generating these synthetic data could be basic (e.g., assuming a nominal resolution and surface brightness limit) but such simplifications will not suffice in the 2020s. 

We outline in \S\,1 the scientific cases for synthetic data and in \S\,2 how realistic synthetic data can mitigate risk for future observatories. As we discuss in \S\,3, both of these cases lead to the conclusion that observatories should be funded to provide robust data simulators for their telescopes and instruments. Moreover, to fully realize the potential that synthetic data provide, both data simulators and synthetic data need to be made publicly available.  These recommendations will help maximize the science return of future instruments and observatories.

\noindent
{\bf \goodblue 1. The scientific need for synthetic data:}
 We discuss here three scientifically-motivated use cases for synthetic data: testing complicated measurement methods (\S\,1.1), making predictions for specific observations (\S\,1.2), and testing physical models of the universe (\S\,1.3). As Figure~1 highlights, each of these requires understanding the translation from a physical model of the universe to the ``intrinsic sky'' that universe provides through to realistically-modeled synthetic data.

\noindent
{\bf \goodblue 1.1. Testing complicated measurement methods:} Nearly all fields of astronomy are seeing commonly used measurement methods becoming more complex in ways that require a detailed understanding of instrumental effects. This shift in understanding how reliable measurement methods are---and the details of how observational effects can bias results---is being reflected by a simultaneous increase in the number and regularity of data challenges across a wide range of astronomical fields (e.g., \citealp{mandelbaum15,bonaldi18,mandell19}).
We discuss here some science cases with increasingly complex measurement methods requiring robust synthetic data.

    \noindent {\em \goodblue Weak lensing and cosmic shear:} 
    Weak gravitational lensing is increasingly becoming a standard tool for precision cosmology, but statistically and robustly measuring the percent-level systematic changes to galaxy shapes requires an intimate knowledge of purely observational effects that can provide a ``signal'' at the same level. 
    Data simulators for testing weak lensing measurement methods must include effects from optics, such as the field-dependence and ellipticity of the PSF, and detector characteristics, such as interpixel capacitance, defects, and non-linearities. 
Synthetic data are also critical to quantifying how imperfect knowledge of the PSF and of the detector biases measurement of the weak lensing signal. A key example of this type of simulator in weak lensing is the community-driven \texttt{GalSim}\footnote{\url{https://github.com/GalSim-developers/GalSim}} \citep{rowe15} package, which provides a toolkit to make flexible astrophysical image simulation possible across multiple instruments. \texttt{GalSim} has enabled incredible advances in image analysis, and forms the basis of many complex calibration tools for major optical surveys like DES, HSC, KiDS, and LSST \citep{suchyta16, zuntz18, lsst18}. Publicly available tools like \texttt{GalSim}, with modules that simulate standard observatory and detector characterstics for instruments like DECam (DES), LSST, and {\em WFIRST} are {\em critical} for weak lensing science, but the current model for developing these tools will not scale to the levels  on which they will be needed in the 2020s.

\noindent	{\em \goodblue Crowded-field photometry:} Detection and measurement of objects in crowded fields depends not only on the properties of individual objects of interest, but also on the realization of objects\linebreak

\noindent
    \includegraphics[width=\textwidth]{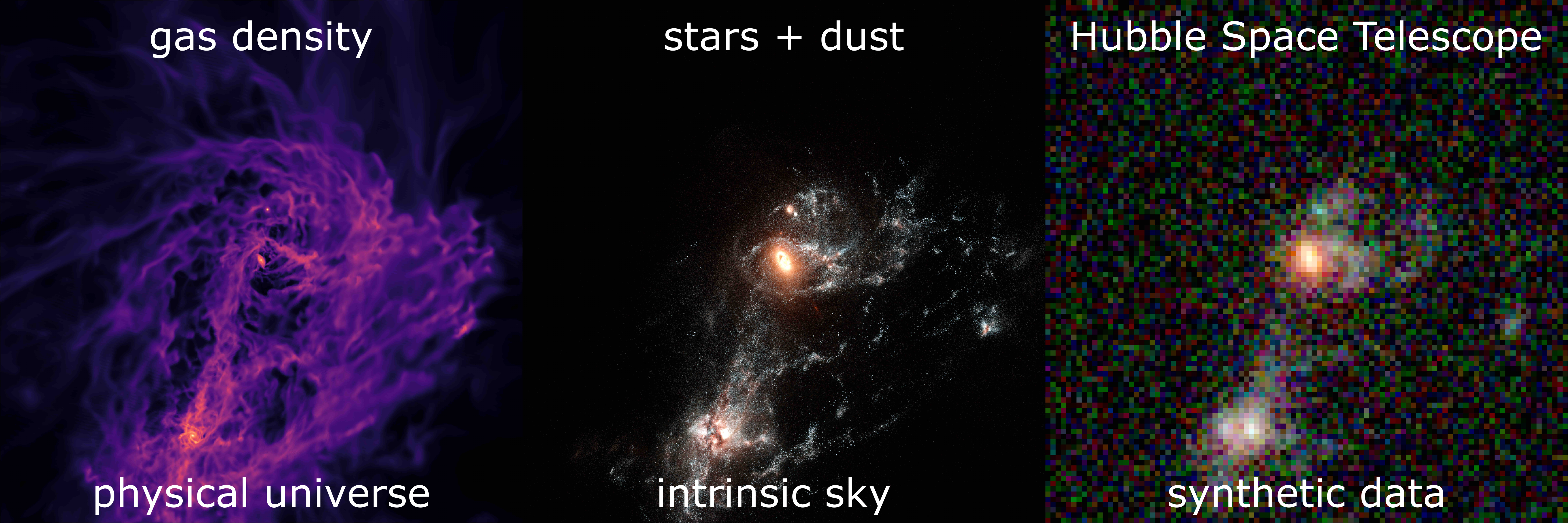}
    Figure~1:
    A cosmological galaxy simulation at $z=2.4$, showing the total gas column density (left), stellar light attenuated by dust (middle), and synthetic image (right) as it might be observed in multi-band {\em Hubble Space Telescope} surveys \citep{simons19}.
    The scientific goal is to be able to use real observations, akin to the synthetic ones in the right panel, to back out information about the physical universe akin to the image in the left panel. While detailed astrophysical models can translate the simulated physical universe (here, the physical properties of gas and stellar populations in a merging galaxy) into the intrinsic sky (here, the distribution of starlight including the effects of dust), the final crucial step of modeling how a real telescope would ``observe'' this system requires also simulating the instrument and optical setup. Generally, the expertise for these two translations lies in different hands; if observatories, however, provide data simulators for their instruments with well-defined APIs, then it will be straightforward for the intrinsic skies such as the one here to be {\em directly} compared to {\em real} data using the {\em same} tools used to analyze real observations.\linebreak

\noindent
crowded around it. Determining detection efficiencies, sensitivity limits, and measurement biases therefore requires detailed injection-recovery simulations in quantities sufficient to ensure that Poisson sampling noise of the simulations does not dominate the error budget. While these simulations are often performed by injecting sources into real data, identical tools can be used in planning or proposing observations to ensure that the scientific goals can be met with the observations requested (see \S\,1.2). Crowded field simulations find wide use in distance estimates using the tip of the red giant branch, star formation histories from resolved stellar populations, and studies of globular clusters, Galactic plane and bulge, and microlensing studies \citep[e.g.,][]{danieli18,nardiello18,surot19,calamida15,mroz19}. To date, many of these types of simulation have dealt with small fields (e.g., {\em HST}), but next-generation instruments will require their application over orders of magnitude larger areas. Common tool sets for injecting sources into real and simulated images, and an archive for storing such simulated data, would improve the efficiency and accuracy of observations and final results.

\noindent	{\em \goodblue  Time-domain astronomy:} Determining the detection efficiency of objects that evolve with time, or move, requires not only  simulated images, but also realistic observing strategies and cadences. 
Realistic time-series of sky conditions (seeing, background, airmass, clouds, etc.) is also important for ground-based surveys. Generating image-level time-series can become prohibitively expensive, so a common strategy is to supplement small image-level simulations with lightcurve-level simulations of larger numbers of objects. On top of the evolving data conditions, time-series simulations must then {\em also} include the time evolution of the objects of interest. These simulations are valuable for nearly all areas of time-domain astronomy, including supernovae and gravitational wave progenitor searches, variable stars, exoplanet searches via nearly every method (transits, radial velocities, astrometry, microlensing, direct imaging), and solar system object searches. Gravitational microlensing surveys, for example {\em WFIRST}'s exoplanet microlensing survey \citep{penny19}, lie at the confluence of challenges facing time domain and crowded field simulations. Next generation facilities (e.g., ZTF, LSST, {\em WFIRST}) add 1--2 orders of magnitude to the scale of the simulation problem over the current largest analyses. As an example of the current state of the art, \citet{mroz19} used ${\sim}250$~CPU years to perform image-level detection efficiency simulations of a small subset of the OGLE microlensing survey; in addition to their larger areas, next-generation surveys will need to perform simulations over higher dimensional parameter spaces.

\noindent	{\em \goodblue Galaxy morphologies:}  No individual simulated galaxy will ever have exactly the same morphology as an individual real galaxy, yet with a systematic comparison of simulated galaxy morphologies and those in real, large galaxy surveys potentially holds clues to the details of how, e.g., star formation quenching and morphological changes are coupled. Central to doing this comparison robustly is the ability to create a simulated survey with similar properties to the real survey in question. 
As galaxy morphologies measured from the real data will depend on the resolution, bandpass, depth, and noise properties in the detector, it is not possible to robustly compare observed morphologies to the ``intrinsic'' ones produced by hydrodynamic simulations (Figure~1). Realistic synthetic data are therefore required to test the extent to which different morphological measures reflect interesting true underlying galaxy properties (such as, e.g., whether or not the galaxy is undergoing a merger).

\noindent	{\em \goodblue  Coronagraphy:} High-contrast direct imaging of exoplanets and disks requires blocking starlight using a complex series of optical masks. These masks create field-dependent PSFs, throughputs, and distortions, and leave residual ``speckles’’ of starlight that can mimic exoplanets. Thus, developing the proper tools for data analysis and retrieval---and simply understanding what
 measurements can be reliably made---requires a detailed understanding of how the coronagraph impacts a realistic astrophysical scene. Mock-data challenges are just starting for the {\em WFIRST} Coronagraph Instrument\footnote{\url{https://www.exoplanetdatachallenge.com}} and have already revealed potentially problematic image artifacts not previously appreciated, like the scatter of exozodiacal light behind the coronagraph’s occulting mask to greater separations, beneficial effects like the planets appearing bright in gibbous phase where the forward-scattering debris disk is faint, and the need for new data analysis tools that can make use of complementary data sets (e.g., radial velocities) to better retrieve the astrophysical properties of imaged exoplanets. By learning these lessons early on we can better plan to mitigate any potential problems and take advantage of opportunities, either through improved instrument design or changes in operations.

\noindent {\em \goodblue High-resolution X-ray Spectroscopy of Extended Sources:} Though {\it Chandra} and {\it XMM-Newton} both possess gratings which provide high-resolution X-ray spectroscopy, observations made with these instruments are necessarily limited to point sources or small, bright objects. The next generation of X-ray observatories ({\it XRISM}\footnote{\url{https://heasarc.gsfc.nasa.gov/docs/xrism/}}, {\it Athena}\footnote{\url{https://www.the-athena-x-ray-observatory.eu/}}, and the proposed {\it Lynx}\footnote{\url{https://www.lynxobservatory.com/}}) will provide high-resolution spectroscopy for extended sources using microcalorimeters, which will enable the detailed measurement of velocities and the composition of hot plasma \citep[such as achieved with the microcalorimeter observations of the Perseus Cluster by][]{hitomi16}. Extended X-ray sources, such as supernova remnants and the hot halos of galaxies and galaxy clusters, are highly susceptible to projection effects (which can particularly complicate high-resolution spectroscopy, e.g., \citealp{zuhone16}), and are also limited by low-count statistics and background. Understanding the impact of these effects requires forward modeling of X-ray event lists, including all instrumental effects, from three-dimensional simulations.

\noindent
{\bf \goodblue 1.2. Predicting observational results based on models and astrophysical simulations:} At some level a goal of all astrophysical models is to predict observational results, but bringing models of the physical universe into an observational parameter space requires data simulators that adequately model observational effects. 
For example, measurements in large surveys, such as are and will be taken with PanSTARRS, LSST, and {\em WFIRST}, regularly require incorporating models of the survey footprint, strategy, and cadence in order to compare the number of detected objects (be they dwarf galaxies or variables such as RR Lyrae) to the ``expected'' number. 
Well-developed and publicly accessible data simulators would allow proposals to include detailed mockups of their proposed observations and results, helping TACs better select the best possible science.  

A working example of this kind of simulator are the tasks used for simulating radio astronomical data that are included in the Common Astronomical Software Applications (CASA\footnote{\url{https://casa.nrao.edu/}}) software package. Given that radio interferometers act as spatial filters, and the instrumental sensitivity for emission at various spatial scales is set by the distance between the antennas, it can be crucial for the proposed observing strategy to run simulations that convert ideal images into observing products for various array configurations. For the Atacama Large Millimeter/submillimeter Array (ALMA), the dedicated simulator tool \texttt{simalma} in CASA can be used to manipulate fits images by re-scaling the dimensions and adjusting the flux scale into Jy/pixel, and subsequently to turn the ideal images into mock data-sets of complex visibilities for any of the available 12m and 7m ALMA array configurations. These simulated visibilities, corrupted with thermal noise from the atmosphere and instrument (and optionally other effects), represent the Fourier components of the sky brightness distribution that would be measured by the interferometer in the chosen antenna configuration. In turn, \texttt{simalma} can use CASA’s imaging task to convert mock visibility-data of different array configurations, optionally complemented with simulated single-dish measurements, back into simulated images. This is a powerful tool for users determine the array configurations that should be proposed in order to achieve be proposed in order to achieve the intended science.

In the X-ray wavebands, synthetic observations have also been essential tools for proposal planning. (On the other hand, for high-energy $\gamma$-rays, e.g., Fermi-LAT, synthetic observations are not yet generally used for proposal planning.) Demonstrating that such observations are feasible requires showing that they will not be limited by low-count statistics or background for faint sources, or will not be susceptible to the effects of pileup for bright ones. Tools such as \texttt{PIMMS} \citep{mukai93} and \texttt{XSPEC} \citep{arnaud96} are limited to simulations of spectra, whereas others like \texttt{MARX} \citep[for {\it Chandra};][]{davis2012} and \texttt{SciSim} \citep[for {\it XMM-Newton},][]{gabriel2005} attempt to provide the fullest possible simulation of their respective instruments. These however, still require the user to specify the astrophysical models themselves as inputs.

These kinds of data simulators have also been customized for specific science cases. For example, PandExo\footnote{\url{https://pandexo.emac.gsfc.nasa.gov/}} \citep{batalha17} has been developed specifically to aid in proposal planning for the characterization of exoplanet atmospheres as seen in transit.
The proliferation of bespoke data simulation tools developed for specific uses demonstrates a community need for generating and analyzing synthetic data. These tools often either have a tradeoff between simplistic assumptions about the instruments and observing modes while incorporating complex underlying astrophysical models (e.g., \texttt{Trident}, \citealp*{hummels17}; \texttt{MYOSOTIS}, \citealp{khorrami19}; or customized inclusions of pixelixation, noise, and PSF, \citealp{barrow17,snyder19}) or detailed models of, e.g., detector characteristics at the expense of potentially less well-modeled astrophysical sources (\texttt{GalSim}, \citealp{rowe15}). {\bf  \goodblue  The current state-of-the-art in data simulation lacks modularity}: well-characterized instrument  simulations are often unable (or only with great difficulty) to be directly plugged into the  complex intrinsic skies astrophysical models produce. Moreover, the instrument simulations developed for specific science cases rather than by the observatories maintaining the physical instruments often go stale, as there is not a direct loop between changes in instrument performance and characterization and the models included  in the data simulators.

\noindent
{\bf \goodblue 1.3. Comparing models and astrophysical simulations to observations:}
A logical consequence of the kinds of model-based synthetic data discussed above is that these data enable analyzing synthetic and real data using the {\em same} tools. The
science cases and complex measurement methods discussed in \S\,1.1 are primarily aimed at answering ``are we measuring what we think we are measuring?'' A powerful appeal of synthetic data generated from astrophysical models, however, is that for the models we {\em know} what the underlying physical universe is: how observations of these mock universes are or are not in agreement with those from the real universe can therefore give insight into the physical properties underlying the real observations. Making these comparisons in the observational plane is the cleanest way to both make as few assumptions about the data as possible and to effectively track which assumptions {\em are} made when required.

\noindent
{\bf \goodblue 2. The engineering need for synthetic data:}
Figure~2 gives a schematic for how instrument simulations and astrophysical simulations can be combined via a ``data simulator'' to produce realistic synthetic data.
The science cases above motivate the need for robust data simulators that can ingest astrophysically interesting ``intrinsic skies'' and output realistic synthetic data. For most of these cases, a mock final data product (e.g., images with flat field corrections applied, or spectra that have been continuum-fitted) may suffice for the that specific scientific goal. However, many of the above cases are examples in which uncalibrated data are of interest {\em because} of the need to test how, e.g., detector imperfections affect measurements. 
From the engineering perspective of preparing for new observatories and instruments, however, the situation is somewhat reversed: the details of the observatory and the instrument specifics may be modeled extremely well, but if these simulators lack the ability to {\em also} incorporate mock astrophysical skies, then much of their potential power will be unrealized.  Critically, by bridging this gap, synthetic data can 
 mitigate risk for future observatories and missions by effectively priming and testing pipelines.

Simulations can demonstrate which observations are likely to pose a risk to instrument operation or health and safety. For example, very bright X-ray and UV sources may not be safely observed at all or may require special handling (e.g., moving the source itself off of the instrument and using diffraction gratings to disperse the spectrum onto the field of view). Such determinations can be made using synthetic observations in a way which does not require repeated testing and destruction of prototype instruments.

    \noindent
    \includegraphics[width=\textwidth]{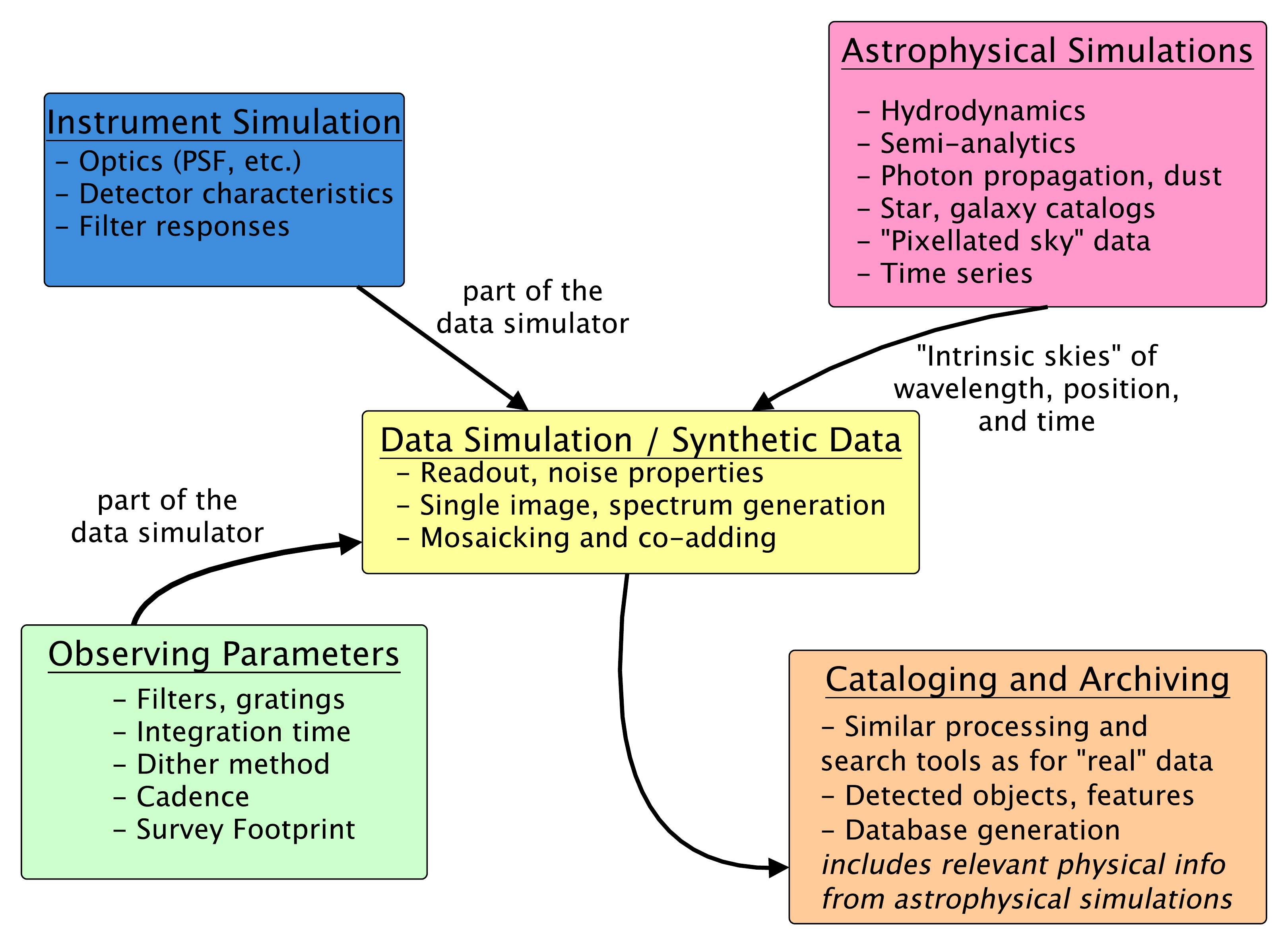}
    Figure~2: Schematic showing how a data simulator consisting of both instrument simulations and observing parameters combine with astrophysical simulations to create synthetic data. These synthetic data would then be able to be ingested into searchable archives while preserving the link back to both the underlying  physical information contained in the astrophysical simulations and the observational information from the data simulator.\\
 
\noindent
{\bf \goodblue 2.1. Lessons learned from {\em JWST}:}
As experience with {\em JWST} has shown, science-free engineering data from the flight hardware is not enough to prime and test the pipelines effectively. In this example, a simulator was provided to aid in the development and testing for the Science \& Operations Center (S\&OC), but it does not go as far as modeling the full complications that real science observations will have. In short, this simulator, the Observatory Testbed Simulator (OTB), attempts to model {\em JWST} in terms of behavior and output products that will be generated once {\em JWST} is on station via a combination of hardware and software components. Some of the hardware components are identical to those that are on {\em JWST}.  The OTB also runs the {\em JWST} flight software, instrument simulators for each of {\em JWST}'s four instruments, and simulates other critical {\em JWST} behavior such as attitude control, proposal execution and telemetry generation. The OTB does a good job simulating these critical {\em JWST} functions and is key to fully exercising the {\em JWST} S\&OC prior to launch.
The OTB is used during S\&OC testing, whereby (mock) proposals are scheduled, uploaded to the telescope and executed.  The corresponding observations generate detector readout commands that store the data onto the solid state recorder (SSR). In reality, during communication contacts, the SSR data, which consists of both engineering and science data, will be downloaded through the Deep Space Network (DSN) and then transferred to the S\&OC, for eventual processing through the science pipelines in the Data Management Subsystem (DMS). 
Unfortunately, the synthetic detector data the OTB presents to DMS consists of only test pattern data. This is insufficient in many areas for high-fidelity exercising of the calibration pipeline software. Most notably lacking is the image alignment of dithers and mosaics, which require a matching overlap of detector data. Likewise, pipelines for source or spectral extractions cannot be fully tested.
Grism, integral field spectroscopic, aperture masking  interferometry, and coronagraphic observations are also unable to be robustly modeled and thus have their relevant pipelines fully tested in an end-to-end manner pre-launch.
Thus full end-to-end processing of realistic science data is not achievable using data from the OTB: stand-alone processing using simulated data is the {\em only} way to fully exercise  the calibration pipelines. 
{\em \goodblue  These needs will only increase in the 2020s, for both ground- and space-based observatories.}

\noindent
{\bf \goodblue 2.2. Lessons from precision weak lensing experiments:}
In particular for the precise weak lensing science cases that drive a large part of the planned cosmological science of the 2020s in surveys like LSST, {\em Euclid}, and {\em WFIRST}, the role of a data simulator is now a fundamental and key part of infrastructure development for both the observatory and survey, and later the pipeline infrastructure for data reduction and analysis. 
Current surveys like DES rely increasingly on full survey-scale simulation efforts that reproduce almost every stage of the data reduction pipeline in order to test and calibrate our weak lensing measurements. 
These simulations rely on public tools like {\texttt GalSim}, but incorporate proprietary knowledge and information from the survey. Examples of this include full suites of image simulations that mimic the DES data to calibrate the shear measurement response matrix \citep{sheldon17, zuntz18} and synthetic injections of object models that are derived from deep-field data templates into single-epoch images that span the survey footprint and timeline in order to calibrate \citep{suchyta16}. 
These are used to calibrate (1) detection efficiency, (2) responses to lensing magnification, (3) completeness of our lensing selection bias estimates, (4) selection function for galaxy samples. 
Re-running the full data processing pipeline from the detection stage forward is required on the images with synthetic injections. 

The LSST DESC has gone further than DES in simulating a large chunk of survey operations from first principles, modelling the observatory and detectors as well as the astrophysical and cosmological simulation being observed to accomplish a similar set of goals,  including providing a testbed for algorithm and pipeline development prior to being run on the real data. 
The {\em WFIRST} high-latitude survey cosmology Science Investigation Team is pursuing a similar path, producing large suites of simulations to inform requirements testing and algorithm development. 
These synthetic data sets are often of similar size to the actual survey data--and sometimes larger---with a computational and person-hour cost on par with the reduction and analysis of the data they calibrate.
Because of the fundamental necessity of these synthetic data sets to enabling the precision weak lensing science promised in the 2020s, it is of great benefit to all experiments to enable and support community wide efforts to develop common simulator tools and archive existing synthetic data sets that have become necessary for supporting the reproducibility of research.

\noindent
{\bf \goodblue 3. What is required to make this happen:} 
Given the above considerations, we have two recommendations for current and upcoming observatories. While both will require additional funding to set up, and a relatively smaller amount of additional funding to maintain, they are both extensions of services observatories already provide: exposure time calculators and data archives.

We note that both of these recommendations could have enormous benefits if implemented very early in the conception and development of new facilities. Often if these funds are allocated, they are not allocated until {\em after} design requirements are set, paradoxically cutting off the ability to make informed simulation-based science-motivated decisions about how to set those requirements. Funding robust data simulators for NASA's next generation of flagships before or in Phase A will not  only go a long way towards optimizing the design of these missions, but will in the long run help keep costs down as  design changes at later stages are expensive \citep{bitten19}.

\noindent
{\bf \goodblue 3.1 Facilities will need to provide {\bf \em and maintain} robust data simulators for their instruments, telescopes, and surveys.} 
From both a funding and a knowledge-base standpoint, the most sensible path forward is for the facilities tasked with maintaining instruments to also be tasked with constructing and maintaining data simulators for their instruments. Not only will the facilities themselves benefit from these tools (\S\,2), but the resulting synthetic data will help maximize the science return from their instruments and surveys. To a certain extent, a natural growth of Exposure Time Calculators (ETCs) that most observatories already provide and maintain, can form the base for more sophisticated simulation software (e.g., the {\em JWST} ETC\footnote{\url{https://jwst.etc.stsci.edu/}}, \texttt{Pandeia}, is built on a basic data simulator; \citealp{pontoppidan16}). A next step could be to tie simulators directly to observing tools at the proposal stage, where all relevant instrumental settings are being specified for the observations. By linking simulators to data from the Virtual Observatory, users can create and verify simulated sky images as part of their time request.

Traditionally, there have not been any requirements (or, indeed, motivation) to provide public data simulators, but we are reaching a point where these tools are {\em necessary} for scientific reproducibility and for fully interpreting astronomical data. When the onus for generating data simulators is placed on the community, they are expensive to make and often require proprietary knowledge to create (or even use). For example, while many of the details for basic data simulators may be publicly available (e.g., PSF Zernike coefficients or detector characteristics), it takes individuals an inordinate amount of time and energy to recast these details into something that can be plugged into customized data simulators (e.g., the ones we discuss in \S\,1.2). 
Moreover, forcing the community to develop  these tools can open the door to avoidable confusion, such as when results disagree owing not to the science at hand but to 
misunderstandings of how the observatory or instrument functions. Facility-provided data simulators provide a chance to enhance scientific productivity by avoiding these unnecessary distractions.
Part of the solution here will be to define community-standard ways of providing this information, but a longer-lasting approach will be for, to the extent possible, data simulators to be modular and share common tools. Such an approach will increase usability and decrease cost.

Coordinated efforts among observatories and instruments will lead to more robust and easier to use tools for simulating observations for a wide range of space- and ground-based facilities. For example, in the radio regime (where many facilities operate under open skies policies), generic simulation tools within the CASA software can in principle be used for a wide range of radio telescopes. For future radio telescopes, such as a Next-Generation Very Large Array \citep{murphy18}, these simulation tools will need to be improved and tied in with more realistic physical sky-models (\S\,2). Nevertheless, they show the potential for developing generic simulators that will be widely used by the astronomical community.

We emphasize  that, in addition to being as modular as possible, these data simulators should be open source and built using a commonly accessible language. Moreover, as many science cases require Monte Carlo type calculations over large datasets that can only be generated with  full end-to-end forward modeling (e.g., many coronagraphic applications), data simulators should be as numerically efficient as possible.

We also note that developing these simulation tools is but one in a long example of the critical role software development will increasingly play in astronomy in the decades to come; we refer the reader to the A.\ M.\ Smith et~al.\ white paper, ``Elevating the Role of Software as a Product of the Research Enterprise'' for more thorough discussions and associated recommendations on this topic.

\noindent
{\bf \goodblue 3.2 Synthetic data will need to be made publicly available in archives (much like real data) so as to lower the barrier of entry to all.}
While the several community efforts to put simulations in publicly-accessible archives, such as the Theoretical Astronomical Observatory (TAO\footnote{\url{https://tao.asvo.org.au/tao/}}), signal a desire in the community to be able to share mock data in the same way as real data, these repositories lack the instrument models and ability to construct synthetic data that is necessary for a {\em true} comparison of theory and reality. By making synthetic data publicly available in archives that are searchable (in the same ways as archives for real data are), observers will no longer be required to have proprietary access to simulations and models for generating predictions and making comparisons to their data. Likewise, if there are well-defined data containers and APIs for the community to package their simulations and models in so that they can be read into data simulators (i.e., to be ingested into synthetic data archives), theorists will no longer be required to have detailed knowledge of specific instruments in order to make relevant predictions for what these instruments will observe (detailed knowledge of instrument performance will undoubtedly always be beneficial to data interpretation). Thus publicly available synthetic data and accessible data simulators will increase science return of both real {\em and} mock data.

A key piece of these synthetic data archives will be the ability to link back to the underlying {\em physical} information in the models from which the synthetic data are generated (Figure~1). The community desire is {\em not} for simply having more mock data to analyze but for data for which the underlying truth is known. Care will therefore need to be taken that the synthetic data metadata (and likely associated auxiliary files or links back to publicly available astrophysical simulations) are not only well-documented and well-maintained but also include information about the astrophysical models, physical properties of the simulated objects, and the resulting intrinsic skies.

We envision two possible approaches, depending on both the scale of the possible synthetic data and the computing resources available to an individual facility or archive. The first approach
would be for archives to ingest the intrinsic skies provided by models and provide back synthetic data produced on-the-fly via user-selected observing modes. While this kind of computation will become less costly in the 2020s, a complementary approach will be to provide only stock examples of fully synthetic data but to enable users to access the intrinsic skies via either download or cloud-based computing and run the data simulator with the desired parameters themselves. Regardless of the specific approach, it will be for many science cases as critical to provide the noise- and instrumentation-free intrinsic skies to the user as it will be to provide the realistic synthetic data.

\pagebreak


\end{document}